\documentclass[lettersize,journal]{IEEEtran}
\IEEEoverridecommandlockouts
\usepackage{cite}
\usepackage{amsmath,amssymb,amsfonts}
\usepackage{algorithmic}
\usepackage{graphicx}
\usepackage{textcomp}
\usepackage{xcolor}
\usepackage{comment}

\usepackage[colorinlistoftodos]{todonotes}    
\usepackage{color,soul}
\usepackage[numbers, sort]{natbib}
\sethlcolor{yellow}
\usepackage[font=normalsize,labelfont=bf]{caption}

\newcommand{\etal}{et al.}
\usepackage{tabularx}
\usepackage{pifont}
\newcommand{\cmark}{\ding{51}}%
\newcommand{\xmark}{\ding{55}}%
\begin{document}

\title{Unified 5G–IoT Framework with CAMARA Gateways and SDN Federation\\
}


\author{\IEEEauthorblockN{Zihan Jia\IEEEauthorrefmark{1},
        Ze Wang\IEEEauthorrefmark{1}, 
		Chen Chen\IEEEauthorrefmark{2}, 
        Ziren Xiao\IEEEauthorrefmark{1} and
		Fung Po Tso\IEEEauthorrefmark{1}}\\
	\IEEEauthorblockA{
		\IEEEauthorrefmark{1}Department of Computer Science,
		Loughborough University, UK\\
		\IEEEauthorrefmark{2}Department of Computer Science and Technology, University of Cambridge, UK\\
		Email: \{z.jia, z.wang6, z.xiao, p.tso\}@lboro.ac.uk;
		cc2181@cam.ac.uk;
	}\vspace{-3ex}}

\maketitle

\begin{abstract}


The convergence of 5G and IoT enables fully connected, intelligent environments, but it faces challenges from the fragmentation of public/private 5G networks and the heterogeneity of IoT networks. We propose a unified framework using CAMARA open gateways, which provide standardized, open APIs to expose network capabilities, reducing fragmentation and simplifying interoperability, supported by a federated SDN architecture that ensures scalable cross-domain control. We further demonstrate 5G-based remote control of KNX devices, extending industrial and building automation. These contributions lay the foundation for a secure, dynamic “network of networks” supporting next-generation applications.

\end{abstract}


\section{Introduction}


The integration of IoT systems with both public and private 5G networks remains a major technical challenge. Several issues hinder this process: (i) fragmentation of protocols, vendors, and infrastructure, which prevents seamless interoperability; (ii) IoT devices often operate in isolation, making it difficult to coordinate across systems; (iii) private and public 5G networks are managed separately, creating silos; and (iv) the lack of interoperability limits unified management, seamless connectivity, and the full exploitation of 5G capabilities.

To address this, our work focuses on leveraging a CAMARA gateway~\cite{camararesources} that bridges public 5G, private 5G, and heterogeneous IoT ecosystems. The open gateway acts as a unified control and data exchange layer, enabling devices and services to operate across network boundaries without compromising performance or security.

In our architecture, public 5G networks provide wide-area access, while private 5G networks ensure localized, high-reliability connectivity for industrial and enterprise use cases. The gateway facilitates device discovery, protocol translation, secure onboarding, and data routing between networks. This approach supports dynamic workload migration, centralized monitoring, and end-to-end orchestration of IoT devices regardless of the underlying network.

By focusing on concrete integration mechanisms, such as secure identity mapping, cross-network session continuity, and real-time policy enforcement, our solution enables practical deployment scenarios like smart manufacturing, cross-site asset tracking, and hybrid 5G-IoT services. This work lays the groundwork for a unified 5G-IoT infrastructure capable of supporting both scalability and interoperability in real-world deployments.

To overcome the above challenges, this work proposes a unified architecture, 5G-IoTNet, that integrates private 5G, public 5G, and IoT connectivity, ensuring secure and seamless communication. The key contributions of this work include:
\begin{itemize}
    \item We design a unified framework that enables seamless integration across private 5G, public 5G, and IoT networks by leveraging the customized CAMARA open gateway architecture.
    \item We develop a federated, SDN-driven architecture to facilitate dynamic and scalable network management across heterogeneous domains.
    \item We demonstrate remote control of KNX devices via 5G connectivity, bridging industrial automation with next-generation mobile networks through a real-world, onsite trial in an operational commercial building.
\end{itemize}

\section{background and related work}
\subsection{Background}

\subsubsection *{CAMARA Open Gateway}
CAMARA Open Gateway\cite{camararesources} is a framework developed by GSMA\cite{gsma_open_gateway_api_descriptions} and major operators that provides standardized APIs to securely expose telecom network capabilities. It abstracts complex network functions, enabling developers to access services like identity verification, device location, and quality control through uniform, RESTful APIs. Using OAuth2 and OpenAPI specs, CAMARA ensures consistent, scalable integration across different operators, making telecom networks as accessible as cloud services.

\subsection{Relate Works}
Federating IoT, private, and public 5G networks enables seamless interoperability, offering benefits such as expanded coverage, mobility, and optimized resource use for enterprises, operators, and end-users. However, as illustrated in Table~\ref{table::related}, most existing work focuses on individual network technologies, overlooking the growing need for comprehensive federation.

Some studies explore public network federation. For instance, Kiela~\etal~\cite{Kiela2021} propose a Network-in-a-Box for federating 4G and 5G standalone systems, achieving high downlink throughput. Bartolín-Arnau~\etal~\cite{BartolinArnau2023} analyze 5G network parameters to meet industrial application requirements.

Other work addresses private 5G configuration. Bektas~\etal~\cite{Bektas2021} and Brown~\etal~\cite{brown2019private} focus on demand-based planning. Wu~\etal~\cite{Wu2022} deploy industrial applications using private 5G with edge cloud and RAN. Kao~\etal~\cite{Kao2021} develop a mobile edge system (5G Intelligent A+) enabling bandwidth management and local breakout.

Federation of IoT networks has also been investigated. MIFaaS~\cite{Farris2017} supports latency-sensitive IoT services via cloud federation. Massonet~\etal~\cite{Massonet2017} enable cross-platform network slicing with unified security.
Finally, Li~\etal~\cite{9496666} propose a multi-domain solution integrating private 5G via public networks. Valcarenghi~\etal~\cite{Valcarenghi2018} present an orchestrator for resource and domain federation in 5G.

In summary, while some efforts explore network federation, few address the unified integration of private 5G, public 5G, and IoT networks, limiting the full realization of their combined benefits.

\begin{table}[t]
\centering
\caption{Summary of Existing Work}
\begin{tabular}{|c|c|c|c|}
\hline
 Work & Public 5G & Private 5G &  IoT\\
\hline
Valcarenghi~\etal~\cite{Valcarenghi2018} & \cmark & \cmark & \xmark \\
Kiela~\cite{Kiela2021} & \cmark &\xmark &\xmark \\

Arnau~\etal~\cite{BartolinArnau2023} & \xmark &\cmark  &  \xmark\\
Bektas~\etal~\cite{Bektas2021} &\xmark &\cmark  &\xmark\\
Brown~\etal~\cite{brown2019private} &\xmark & \cmark&\cmark \\
Wu~\etal~\cite{Wu2022} & \cmark  & \cmark&\xmark \\
Kao~\etal~\cite{Kao2021} &\cmark  & \cmark&\xmark \\
MIFaaS~\cite{Farris2017} & \xmark & \xmark & \cmark\\
Li~\etal~\cite{9496666} &\xmark  & \cmark& \xmark \\
Our work &\cmark  & \cmark& \cmark \\
\hline
\end{tabular}
\label{table::related}
\end{table}

\section{System Design \& Architecture}


\subsection{Overview}



\begin{figure}[t] 
    \centering
    \includegraphics[width=0.5\textwidth]{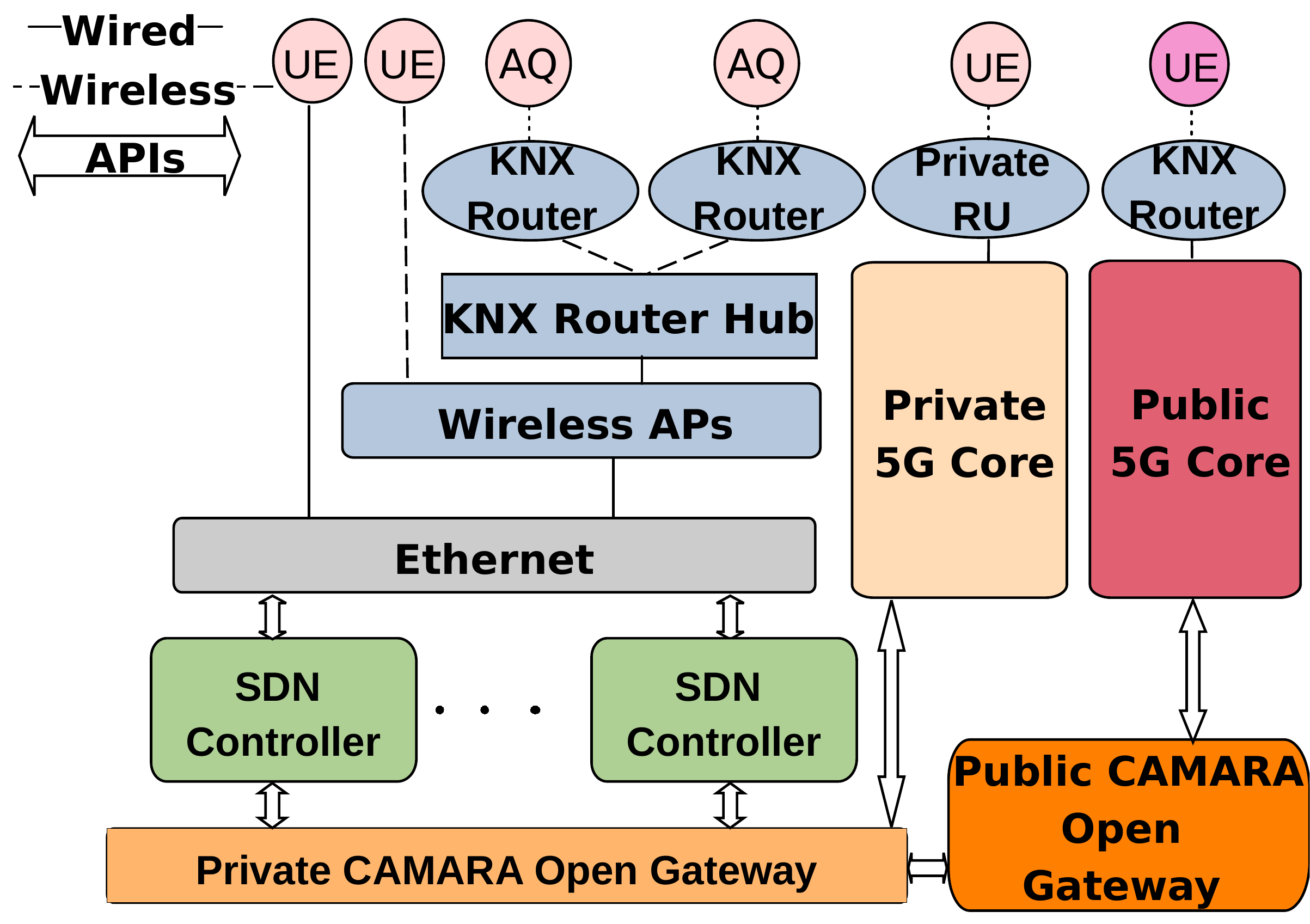} 
    \caption{5G-IoT system architecture} 
    \label{fig:sys} 
\end{figure}

The proposed architecture, as shown in Figure~\ref{fig:sys}, enables secure, scalable, interoperable communication by integrating private and public 5G networks via standardized CAMARA Open Gateway APIs, coordinated through a centralized SDN controller. This hybrid design combines deterministic private 5G control with policy-driven public network interaction, ensuring both local reliability and secure service extension.

Each SDN controller manages its own domain but participates in cross-domain orchestration, coordinating Open Gateway interfaces to enable dynamic policy enforcement, traffic steering, and service differentiation. This distributed control is essential for scalability.

By addressing the limitations of public-only (lacking local control/QoS) and private-only (isolated) deployments, the system enables seamless, secure integration across network boundaries, supporting demanding 5G applications.

\subsection{Private 5G Network and Public 5G Network Domain}


The system centers on a private 5G network comprising on-site Radio Units, a 5G Core, and enterprise-managed UEs. Fully owned and operated locally, it enables deterministic performance via customizable RAN and core settings. Functions like authentication, traffic control, and network slicing are localized, ensuring precise control over bandwidth, latency, and security. Slicing isolates critical traffic (e.g., industrial IoT, real-time monitoring) to guarantee QoS. Complementing this, the public 5G network, managed by the MNO, offers broad mobility and coverage. Though isolated from private traffic for privacy, it supports coordination via standard APIs. Dual-SIM devices can seamlessly switch between networks.

\subsection{Private \& Public CAMARA Open Gateway Integration}

The implementation of the CAMARA Open Gateway framework in our system is designed to achieve secure, fine-grained, and federated access control across private and public 5G domains. Specifically, the following mechanisms have been realized:

\textbf{1) User Authentication and Onboarding in Private 5G:}

In private 5G deployments, the core network (5G core) typically handles user registration and session management. To augment security and enhance policy flexibility, our system introduces the Open Gateway API as an intermediary authorization agent that actively mediates user access requests. When a user initiates an access attempt toward the private 5G network, the Open Gateway API queries the 5G core to retrieve detailed user credentials and contextual metadata, such as subscription status, device type, and security posture. These data points are then evaluated against a set of predefined, dynamically configurable authorization policies hosted within the Open Gateway framework. If the user satisfies all authorization criteria, the Open Gateway issues a secure command back to the 5G core, instructing it to authenticate the user and instantiate the necessary network slices and resources for their session.
This approach ensures that unauthorized or potentially compromised users are blocked at the gateway layer before reaching sensitive core functions. By abstracting the authorization logic outside the core network, the system enhances security isolation, reduces the attack surface, and facilitates policy updates without disrupting core network stability.

\textbf{2) Identity-Aware Routing and Service Isolation:}

Beyond initial authentication, the Open Gateway API plays a critical role in enforcing identity-aware routing policies within the private 5G domain. Recognizing that users may belong to diverse roles, such as internal employees, third-party contractors, or industrial IoT devices, the gateway dynamically configures routing rules to direct traffic flows accordingly. For example, employees accessing corporate applications may be routed through dedicated, high-priority network slices with stringent QoS guarantees, while IoT sensors transmitting telemetry data may be assigned to isolated network segments optimized for low bandwidth but high reliability. This routing differentiation is implemented through programmable policy rules managed by the Open Gateway, enabling rapid adaptation to changing user profiles or service requirements. Such fine-grained service isolation not only ensures efficient resource utilization but also enhances security by preventing lateral movement between critical network segments, thereby mitigating risks associated with insider threats or compromised endpoints.

\textbf{3) Federated Access Between Private and Public 5G:}

One of the core innovations of the proposed architecture is the federation capability enabled through CAMARA Open Gateway APIs, which facilitates seamless and secure interoperability between private and public 5G networks. A secure, encrypted communication channel is jointly established between the Open Gateways in the private and public 5G domains, enabling real-time exchange of authenticated user identities, authorization contexts, and policy enforcement signals. This design enables users equipped with dual-SIM devices to traverse between private and public networks transparently, maintaining consistent access permissions, session continuity, and security policies. Importantly, this federation model abstracts away direct signaling interactions between the private and public 5G cores, thereby preserving network segmentation and minimizing exposure of sensitive core interfaces to external entities. By confining cross-domain coordination to the Open Gateway layer, the system achieves a balance between flexible user mobility and robust security governance, supporting multi-tenant collaboration scenarios where multiple organizations jointly manage a federated network ecosystem.

Collectively, these implementation strategies position the CAMARA Open Gateway framework as a pivotal component that not only secures and governs user access within isolated 5G environments but also enables scalable federation and policy harmonization across heterogeneous network domains. This layered approach to access control and federation exemplifies a forward-looking architecture that aligns with emerging requirements in enterprise and industrial 5G deployments, where security, scalability, and interoperability are paramount.

\subsection{Federation SDN Controller}

At the system level, a Federation Software-Defined Networking (SDN) Controller is introduced to fundamentally redefine the management paradigm of CAMARA Open Gateways. Traditionally, control over these gateways is confined to a single administrative entity. The proposed federation SDN framework extends this model by enabling multiple independent organizations to exercise shared control, coordinate operational responsibilities, and collaboratively manage distributed Open Gateway instances spanning private and public 5G domains.

The Federation SDN Controller interfaces directly with CAMARA Open Gateways to specify and enforce routing policies, access control mechanisms, and traffic management strategies. Through federated governance and distributed policy synchronization, the controller transforms the underlying network infrastructure into a flexible and extensible service plane. This architecture supports dynamic adaptation to operational variations and facilitates scalable expansion across multiple administrative boundaries, while ensuring consistent and unified policy enforcement.


\begin{figure*}[t] 
    \centering
    \includegraphics[width=0.8\textwidth]{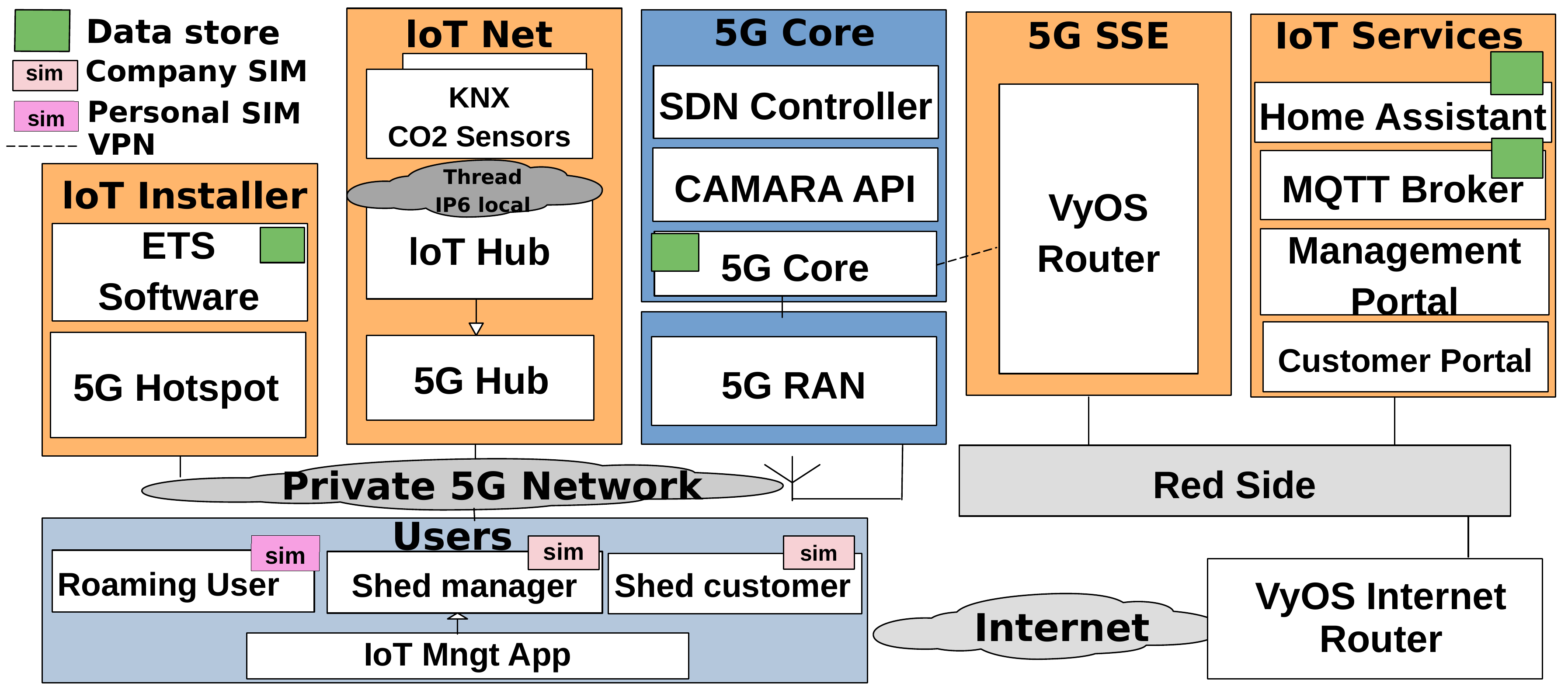} 
    \caption{5G-IoT System Implementation and Deployment} 
    \label{fig:sys_implement} 
\end{figure*}

\section{Use Case}

Intelligent building management needs a system that can monitor and control key subsystems,like environmental sensors, security devices, and energy systems in real time. The system must let different devices and users communicate quickly, reliably, and securely. For example, it should adjust HVAC (Heating, Ventilation, and Air Conditioning) settings automatically based on sensor data, or trigger security alarms when unauthorized access is detected, with minimal delay. 


This requires a control loop that is fast and reliable. At the same time, the system must allow authorized external users, such as mobile staff, remote management services, or third-party contractors, to access certain building functions safely. This requires a flexible and secure interface for controlled collaboration beyond the local network.


With IoT devices spread across the building and 5G networks providing high-speed, low-latency connectivity, the system can maintain smooth communication and continuous monitoring, whether users are inside or outside the building. By combining local, deterministic control with open, programmable access for trusted external actors, the building management system can respond quickly to changing conditions, allocate resources efficiently, and enforce fine-grained policies across all connected devices.

\section{Implementation \& Deployment}

The implementation environment integrates a private 5G network with IoT devices and services, deployed on the Shed testbed~\cite{shed2025}, which is an operational commercial building used for validating real-world scenarios(Fig~\ref{fig:sys_implement}). The IoT network includes KNX devices for lighting, HVAC, and energy management, as well as CO\textsubscript{2} sensors for environmental monitoring. These devices are connected through a local Thread/IPv6 mesh network to an IoT Hub, which aggregates sensor data and control signals and relays them to a 5G Hub for seamless network connectivity. To simplify deployment, the IoT Installer uses ETS software~\cite{knx_ets_home}, the official KNX Engineering Tool for discovering, configuring, and managing smart devices. Through ETS, installers assign group addresses, define device parameters, and establish communication links. The configuration process is supported by a portable 5G hotspot, enabling on-site provisioning, firmware updates, and secure registration of devices to the private 5G network.

The 5G core network runs alongside an SDN controller and the CAMARA API. In our solution, the SDN acts as the top-level controller, while the API continuously reads user-specific information from the 5G core, such as IMSI, IP addresses, and other session details, and passes it to the SDN. The SDN then issues instructions back to the API, which configures the VyOS router by updating its IP routing table. This enables precise traffic control, allowing different users to access only the services they are authorized for. The 5G RAN connects to the private network, providing reliable, low-latency communication for both local and roaming users. Meanwhile, the 5G SSE segment, implemented on the VyOS router, enforces these routing and access policies, controlling both access to specified IoT services within the private network and connections to the red side, which represents the external untrusted Internet.

IoT services such as Home Assistant, the MQTT Broker, the Management Portal, and the Customer Portal are deployed to support building management and user interactions. Access to these services is controlled through the SDN and CAMARA API mentioned above, ensuring that different user types, such as Shed managers, local users, roaming users, and customers, can only reach the services they are authorized for. Users connect using either company or personal SIMs, with VPN connections providing secure remote access when needed.

This setup allows testing of secure, role-based access control, dynamic traffic management, and service orchestration across heterogeneous IoT and 5G components. The combination of standardized Open Gateway APIs, SDN control, and federated policies enables scalable, secure, and interoperable operation in a real-world building environment.

\section{Conclusion}
This paper presents a unified framework that bridges public and private 5G with heterogeneous IoT networks through CAMARA gateways and a federated SDN architecture. By enabling 5G-based control of KNX devices and enhancing security with VPN integration, the proposed approach lays the foundation for scalable, secure, and interoperable connectivity to support next-generation intelligent applications.

{\footnotesize
\bibliographystyle{IEEEtran}
\bibliography{ref}
}
\end{document}